\documentclass[fleqn,10pt]{wlscirep}
\usepackage[utf8]{inputenc}
\usepackage[T1]{fontenc}
\usepackage[T1]{fontenc}
\usepackage{longtable}
\usepackage{booktabs} 
\usepackage{multirow} 
\usepackage{tikz}
\newcommand*\circled[1]{%
  \tikz[baseline=(char.base)]{
    \node[shape=circle,draw,inner sep=1pt] (char) {#1};}}
\title{Data-driven solar forecasting enables near-optimal economic decisions}

\author[1,+,*]{Zhixiang Dai}
\author[2,3+]{ Minghao Yin}
\author[4,+,*]{ Xuanhong Chen}
\author[1]{ Alberto Carpentieri}
\author[1]{ Jussi Leinonen}
\author[1]{ Boris Bonev}
\author[2,3]{ Chengzhe Zhong}
\author[1]{ Thorsten Kurth}
\author[1]{ Jingan Sun}
\author[1]{ Ram Cherukuri}
\author[1]{ Yuzhou Zhang}
\author[1]{ Ruihua Zhang}
\author[1]{ Farah Hariri}
\author[5,3]{ Xiaodong Ding}
\author[4]{ Chuanxiang Zhu}
\author[6]{ Dake Zhang}
\author[2]{ Yaodan Cui}
\author[2]{ Yuxi Lu}
\author[2]{ Yue Song}
\author[2]{ Bin He}
\author[2]{ Jie Chen}
\author[7,8]{ Yixin Zhu}
\author[8]{ Chenheng Xu}
\author[9]{ Maofeng Liu}
\author[10,11]{ Zeyi Niu}
\author[12]{ Wanpeng Qi}
\author[13]{ Xu Shan}
\author[14]{ Siyuan Xian}
\author[14,15]{ Ning Lin}
\author[2,3,*]{ Kairui Feng}

\affil[1]{NVIDIA Corporation, Santa Clara, CA, USA}
\affil[2]{State Key Laboratory of Autonomous Intelligent Unmanned Systems, Tongji University, Shanghai, China}
\affil[3]{Shanghai Innovation Institution, Shanghai, China}
\affil[4]{School of Electronic Information and Electrical Engineering, Shanghai Jiao Tong University, Shanghai, China}
\affil[5]{University of Shanghai for Science and Technology, Shanghai, China}
\affil[6]{Antai College of Economics and Management, Shanghai Jiao Tong University, Shanghai, China}
\affil[7]{School of Psychological and Cognitive Sciences, Peking University}
\affil[8]{Institute for Artificial Intelligence, Peking University}
\affil[9]{Department of Atmospheric and Oceanic Sciences, School of Physics, Peking University}
\affil[10]{Shanghai Typhoon Institute, Shanghai, China}
\affil[11]{Key Laboratory of Numerical Modeling for Tropical Cyclone of the China Meteorological Administration, Shanghai, China}
\affil[12]{Qinghai Meteorological Bureau, Qinghai, China}

\affil[13]{Max Planck Institute for Biogeochemistry, Jena, Germany
}

\affil[14]{Department of Civil and Environmental Engineering, Princeton University, Princeton, USA}
\affil[15]{Andlinger Center for Energy and the Environment, Princeton University, Princeton, USA}
\affil[*]{Corresponding authors:Zhixiang Dai (zhixiangd@nvidia.com),Xuanhong Chen (chenxuanhong@sjtu.edu.cn),Kairui Feng (kelvinfkr@tongji.edu.cn)}
\affil[+]{These authors contributed equally to this work.}

\begin{abstract}
Solar energy adoption is critical to achieving net-zero emissions. However, it remains difficult for  many industrial and commercial actors to decide on whether they should adopt distributed
solar–battery systems, which is largely due to the unavailability of fast, low-cost, and
high-resolution irradiance forecasts. Here, we present SunCastNet, a lightweight data-driven forecasting
system that provides 0.05°, 10-minute resolution predictions of surface solar radiation downwards (SSRD) up to 7 days ahead. SunCastNet, coupled with reinforcement learning (RL) for battery scheduling, reduces operational regret by 76–93\% compared to robust decision making (RDM). In 25-year investment backtests, it enables up to five of ten high-emitting industrial sectors per region to cross the commercial viability threshold of 12\% Internal Rate of Return (IRR). These results show that high-resolution, long-horizon solar forecasts can directly translate into measurable economic gains, supporting near-optimal energy operations and accelerating renewable deployment.
\end{abstract}

\begin{document}

\flushbottom
\maketitle
%
%
\thispagestyle{empty}


\section*{Introduction}

The global energy system is moving toward carbon neutrality, with solar photovoltaics (PV) emerging as one of the fastest-growing renewable technologies~\cite{davis2018net,soergel_sustainable_2021,wang_global_2025}. China aims to raise its solar generation penetration to more than 40\% by 2050~\cite{lu_combined_2021,yan_city-level_2019}. Reaching such an ambitious  target will depend both on continued capacity expansion, and on the ability to manage the inherent variability of solar resources at the consumer level~\cite{yin_impacts_2020,li_global_2020,jerez_impact_2015,huang2020renewable}. For industrial consumers, solar forecast quality determines their daily battery operation and grid interactions, and thus drives their long-term decisions to invest in PV projects~\cite{sobri_solar_2018,ahmed_review_2020,das_forecasting_2018,miller_short-term_2018, coville2025quality}. Industrial consumers must plan battery operations days in advance to remain profitable under peak–valley on-grid electricity price, which requires long-horizon, high-resolution solar forecasts that numerical weather prediction (NWP) often cannot provide~\cite{cui_policy-driven_2025,corwin_solar_2025}.


Recent AI-based weather models such as FourCastNet~\cite{pathak2022fourcastnet}, GraphCast~\cite{lam_learning_2023}, and Pangu-Weather~\cite{bi_accurate_2023} now achieve forecast skills comparable to, or even surpassing, NWP at global 0.25° and hourly scales~\cite{rasp2024weatherbench2,chen_fengwu_2023,sun_data--forecast_2025,price_probabilistic_2025,kochkov_neural_2024,nguyen2023climax}. Many researchers are now drawing findings from low-resolution research to develop high-resolution weather forecasting~\cite{dueben2022challenges,craig2022overcoming,bai2025solarseer}, and therefore to address a wide range of downstream management issues. The transition from low- to high-resolution weather forecasting faces significant challenges, including exponentially increasing computational demands, lack of high-resolution observations, and the cumulation of errors~\cite{reichstein_deep_2019,schultz2021can}. Diverse approaches, ranging from task-specific fine-tuning~\cite{lehmann2025finetuning}, diagnostic modules~\cite{carpentieri_data-driven_2024,xia_accurate_2024}, and multi-source data integration~\cite{bai2025solarseer}, are now being explored. Nevertheless, these advances remain insufficient for industrial consumers, who require forecasts with week-ahead horizons and station-level accuracy~\cite{bodnar_foundation_2025,allen_end--end_2025,benboualleguea2024rise}.

\begin{figure*}[t]
  \centering
  \includegraphics[width=0.8\textwidth]{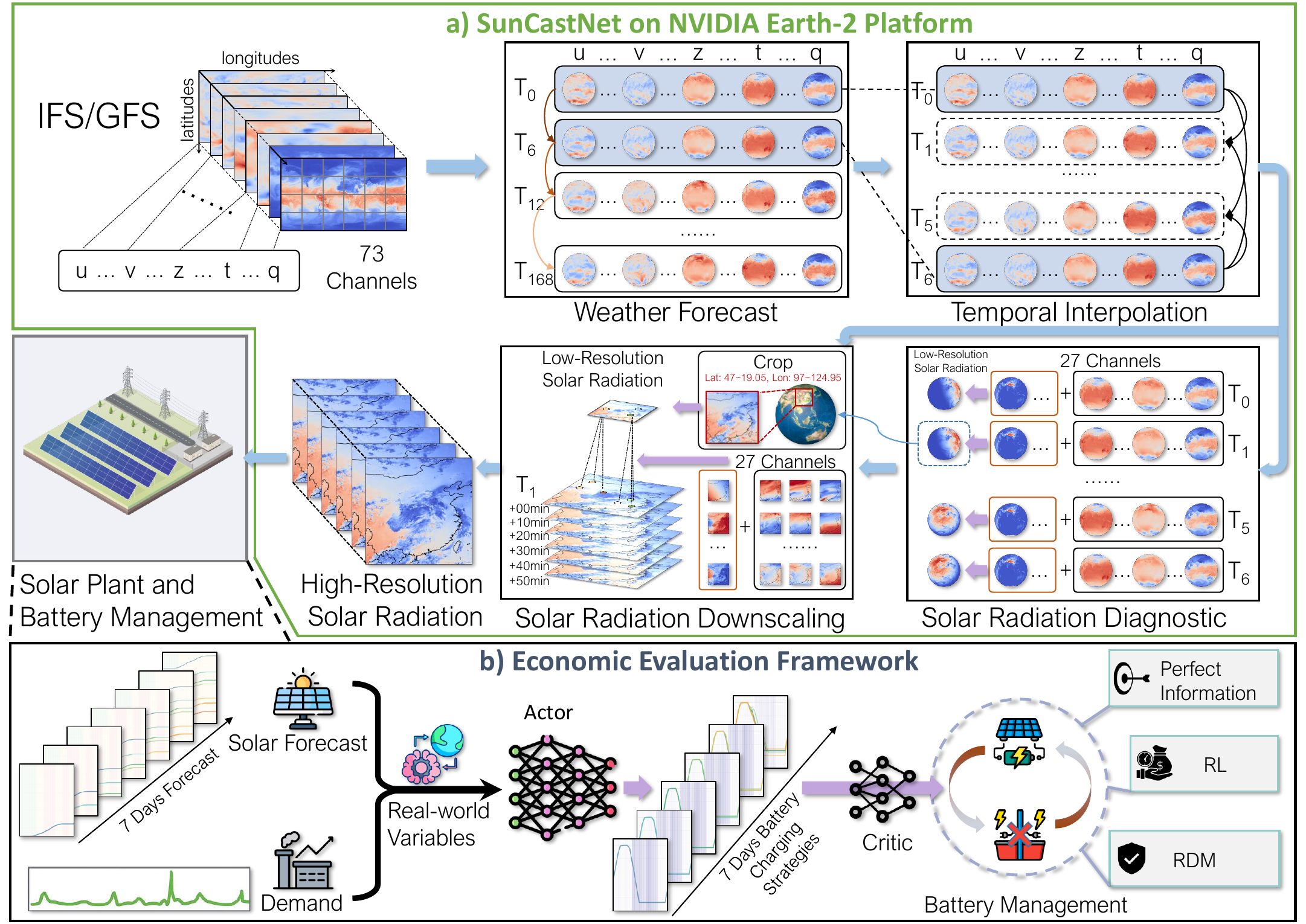}
\caption{\textbf{SunCastNet solar forecasting pipeline coupled with an RL framework for industrial economic evaluation.} 
(a) SunCastNet Framework: The system begins with IFS/GFS (73 atmospheric variables at 0.25°, 6-hour intervals) , processed by a four-stage sequence(a SunCastNet): 
(i) \emph{Weather forecasting} with SFNO, which predicts global circulation using 73 input variables to produce 6-hourly fields at 0.25° resolution; 
(ii) \emph{Temporal interpolation} with ModAFNO, which refines the coarse forecasts from 6-hourly to 1-hourly resolution using two consecutive atmospheric states (2 × 73 variables) together with 9 auxiliary fields (total 155 input channels → 73 output channels);
(iii) \emph{Solar radiation diagnostics} with AFNO, which maps 31 key atmospheric fields to 1-hourly surface solar radiation downwards (SSRD); and 
(iv) \emph{downscaling} with CorrDiffSolar, which transforms 57-channel inputs into 0.05°, 10-minute SSRD fields calibrated against dense East Asia–Pacific observations. 
(b) Economic Evaluation Framework: These high-resolution forecasts are then embedded in RL–based battery management models that integrate solar generation, electricity demand, and price signals to derive optimal charging and discharging strategies. 
By comparing against perfect-information and robust decision making (RDM) baselines, the framework quantifies the impact of forecast skill on operational regret, infrastructure sizing, and long-term investment returns across industrial sectors. 
The analysis focuses on China, covering the domain 47--19.05$^{\circ}$N, 97--124.95$^{\circ}$E.}

  \label{fig:pipeline}
\end{figure*}

Here, we introduce \textbf{SunCastNet} (developed on the NVIDIA Earth-2 Platform), a sequential framework that translates recent advances in AI weather forecast into high-resolution (0.05°, 10-min) long-horizon (7-day) solar forecasts and downstream decision support (Fig.~\ref{fig:pipeline}a). The forecasting component is organized into four successive stages reflecting atmospheric processes~\cite{stengel2020adversarial,hess2022physically}: (i) a Spherical Fourier Neural Operator (SFNO) that models global circulation at 0.25° every 6 hours~\cite{bonev_spherical_2023}, (ii) a Modulated Adaptive Fourier Neural Operator  (ModAFNO) that interpolates coarse six-hourly states to hourly variability~\cite{leinonen_modulated_2024}, (iii) an AFNO-based diagnostic~\cite{carpentieri_data-driven_2024} that maps key atmospheric fields directly to surface solar radiation downwards (SSRD), which is essential to photovoltaic power output~\cite{antonanzas_review_2016} at 0.25° every hour, and (iv) a CorrDiffSolar module that downscales SSRD in (iii) to 0.05° and 10-minute intervals~\cite{mardani_residual_2025} (see the materials and methods), benchmarked against high resolution SSRD data~\cite{letu2022new}. Together, this architecture generates continuous 7-day forecasts validated against 2,164 meteorological stations across China. Compared to Global Forecast System (GFS), SunCastNet achieves 5-10\% lower relative errors, 20\% higher mutual information, and improved forecast consistency.  It takes SunCastNet about 25 minutes to execute every 7-day weather forecast on a single NVIDIA A100 GPU, with an estimated cost of approximately \$0.5 per continental-scale forecast (expressed in 2025 USD). Its training and inference costs remain far lower than those of foundation-model fine-tuning~\cite{lehmann2025finetuning}, satellite-to-forecast end-to-end approaches~\cite{bai2025solarseer}, and traditional NWP~\cite{kalnay1996ncep}.

However, improved forecasting metrics (e.g., RMSE) alone do not necessarily translate into economic value~\cite{mcgovern2017using}. A solar prediction paradox arises because “sunny” conditions are the most common in many regions: a naïve model that always forecasts “sunny” may achieve about 85\% accuracy, yet it would still cause significant losses for battery operators who fail to precharge before cloudy days. This asymmetry explains why forecast "accuracy" can be sometimes misleading~\cite{winkler1994evaluating,jose2017percentage,visser2024probabilistic,campos2022assessing}. When forecasts cannot reliably detect these critical cloudy periods, operators need to resort to robust decision making (RDM)~\cite{bertsimas2011theory,ben2013robust,kim2024sample}, a group of uncertain set-driven methods that maintain defensive reserves to minimize maximum potential regret. Only when forecasts carry more informed content can the stochastic optimization driven methods, such as reinforcement learning (RL)~\cite{jiang2015optimal,brown2025unit}, be viable to optimistically seek average gains.

Here, we evaluate SunCastNet's operational and economic value (Fig.~\ref{fig:pipeline}b) comparing to GFS. To enable the 25-year economic backtesting, we first generated 7-day forecasts at 10-minute resolution with 6-hour issuance intervals throughout the 25-year period, which required approximately 15,000 A100 GPU hours and produced about 43 TB of data. For a given solar forecast , RL agents learn optimal charge–discharge strategies by integrating configuration of solar panels and batteries, demands, and price signals (see the Materials and Methods).  These short-term strategies are applied to 25-year investment backtests across ten industrial sectors where alternative capacity configurations are systematically explored. The results are explicitly advantageous over GFS: reducing decision regret compared to RDM in battery scheduling by 72-93\% (versus 43–66\% for GFS; $50\%\pm25\%$ quantiles), and shifting multiple representative solar projects from “infeasible” to “profitable” in long-term investment analysis. Our analysis also shows that forecast horizon length is pivotal for realizing economic value: a 2-day horizon only reduces regret by less than 40\%, while a 7-day horizon reduces regret by over 70\% in many regions, demonstrating that extended horizons substantially enhance system benefits. These findings suggest that the combination of higher spatial (0.05° vs 0.25°), finer temporal (10-min vs 1-hour) resolution, and longer forecast horizons fundamentally compounds benefits for the economics of industrial solar adoption.

\section*{Results}

\subsection*{SunCastNet Forecast Ability }

\begin{figure*}[t]
  \centering
  \includegraphics[width=0.7\textwidth]{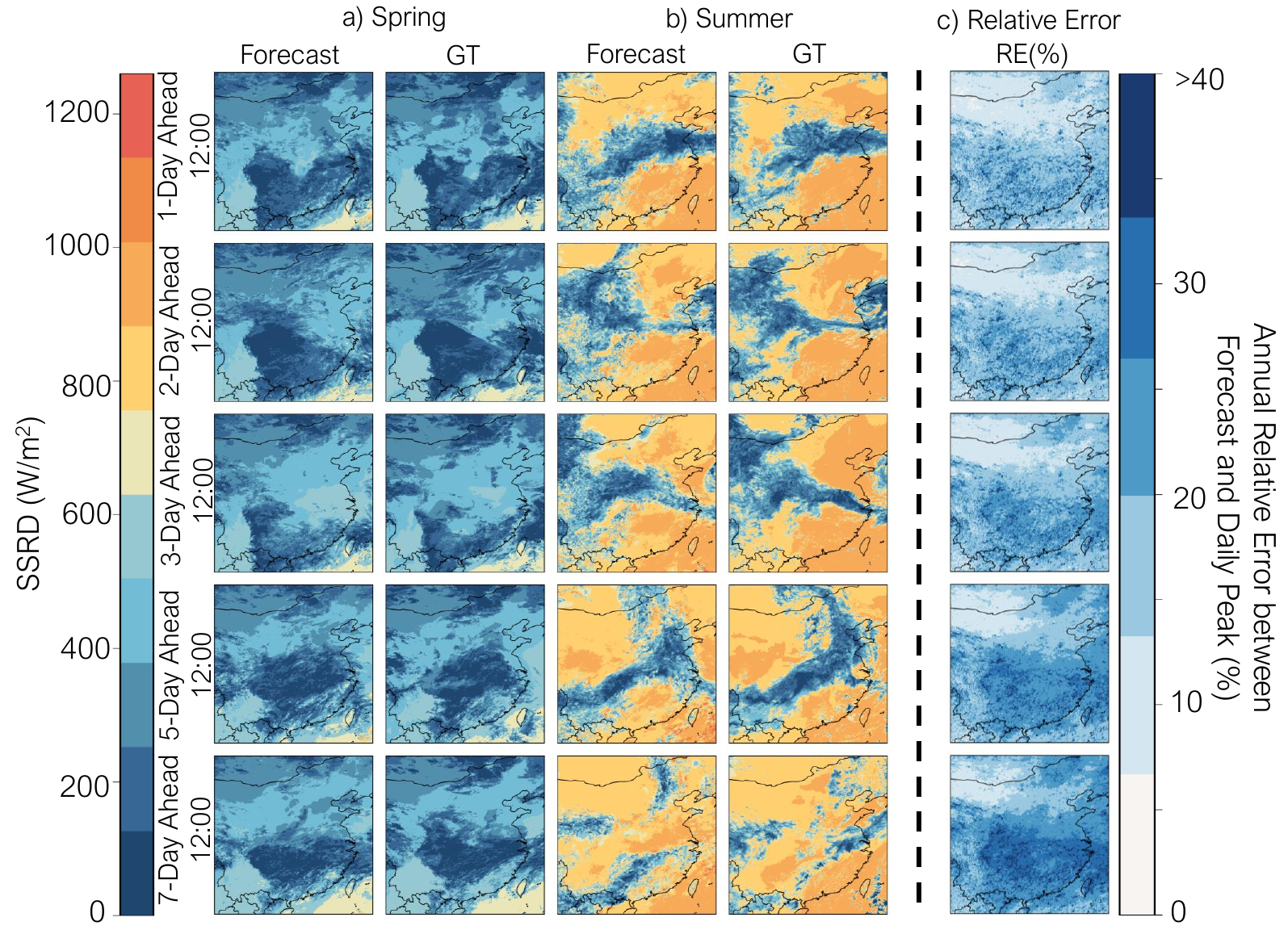}
  \caption{\textbf{Forecast skill of SSRD across seasons and lead times with SunCastNet.} 
  (a) Forecasts at 5-km resolution for a typical spring day (17 January 2020), showing SSRD at 12:00 local time (24-hour clock) predicted from forecasts issued at 03:00 with lead times of 1, 2, 3, 5, and 7 days, compared against satellite-derived ground truth (GT).  
  (b) Same as (a) but for a typical summer day (22 July 2020).  
  (c) Forecast skill as a function of lead time, expressed as the annual relative error (RE) of daily peak SSRD at 12:00 local time (24-hour clock); shading denotes the inter-location error level.}
  \label{fig:forecast_accuracy}
\end{figure*}

\begin{figure*}[t]
  \centering
  \includegraphics[width=\textwidth]{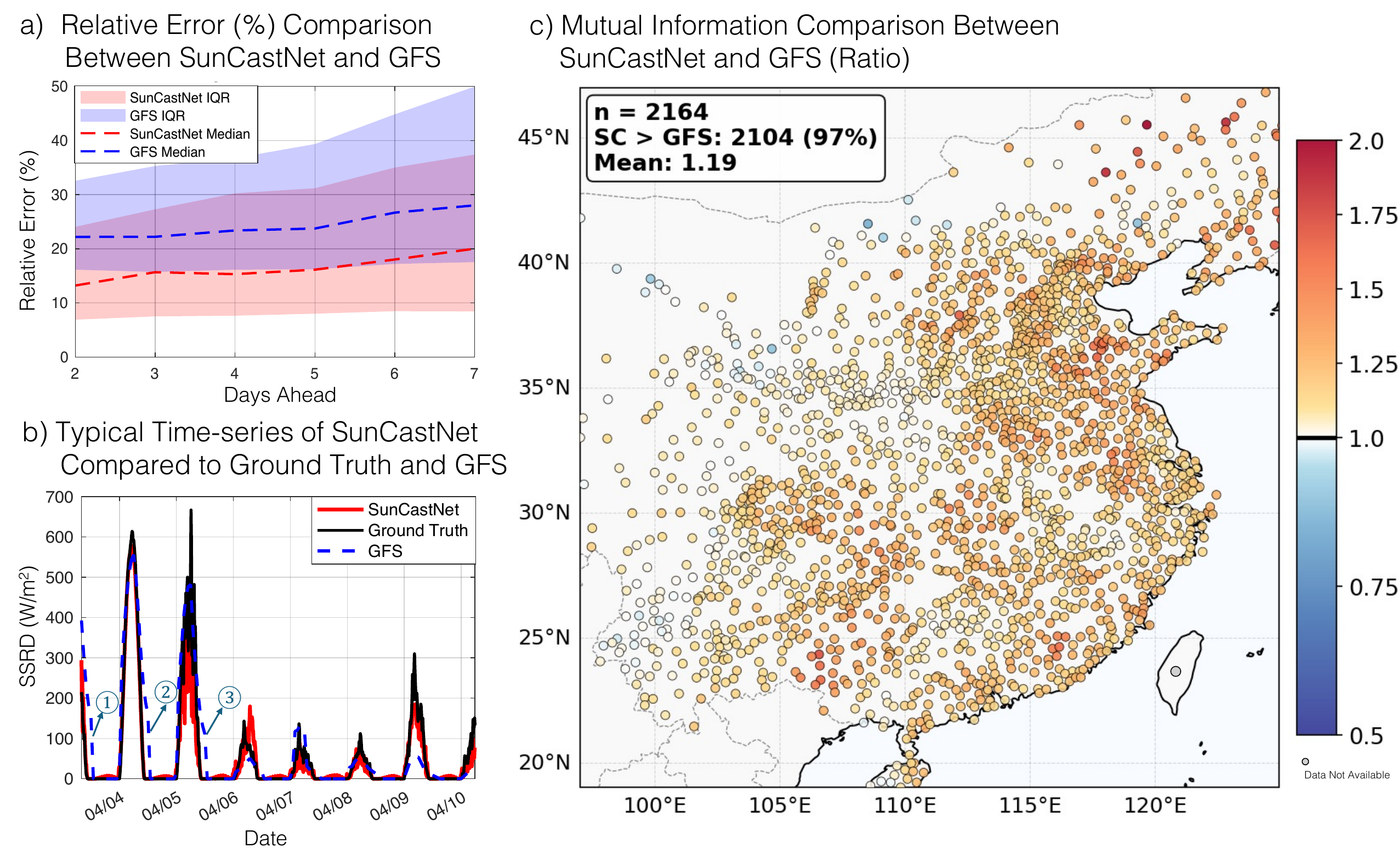}
  \caption{\textbf{Comparison of SunCastNet and GFS forecasts of SSRD.} 
  (a) Relative error as a function of forecast lead time (2–7 days) of daily irradiation across 2,164 stations in China. Red line and shading denote the median and interquartile range (IQR, 25th–75th percentile) of SunCastNet errors, while blue line and shading denote the corresponding values for GFS.  
  (b) Example time series of SSRD at one station (located at the purple star in part c) during early April 2023. Black line shows ground truth measurements, red line shows SunCastNet forecasts, and blue dashed line shows GFS forecasts.  
  (c) Ratio of mutual information between forecasts and ground truth, with colors indicating the station-wise ratio of SunCastNet to GFS; orange/red denotes higher ratios (>1.0) where SunCastNet exceeds GFS, blue denotes ratios <1.0.}
  \label{fig:gfs_comparison}
\end{figure*}

With the SunCastNet pipeline in place (Fig.~\ref{fig:pipeline}), we first examine its core forecasting capability on SSRD.
Figure~\ref{fig:forecast_accuracy} evaluates the forecast skill of SunCastNet for SSRD across multiple lead times and seasons. Panel~\ref{fig:forecast_accuracy}a shows forecasts at 5-km resolution for a typical spring day (16 January 2020). The maps illustrate SSRD at 12:00 local time predicted by forecasts issued at 03:00, with lead times of 1, 2, 3, 5, and 7 days. The SunCastNet forecasts reproduce the broad spatial gradients of solar irradiance evident in the satellite-derived ground truth (GT), including suppressed irradiance over southern China associated with cloud cover. While forecast biases gradually emerge with longer lead times, the overall spatial correspondence with GT remains robust even at a 7-day horizon. Panel~\ref{fig:forecast_accuracy}b presents the same experiment for a typical summer day (20 July 2020), where SunCastNet captures the high-irradiance regions over northern China and the strong cloud-induced gradients across the Yangtze River basin. Similar to the aforementioned spring-day forecast, local cloud structures are distorted slightly, but the large-scale patterns are consistent. This highlights the model’s congruency in its seasonal forecast skill despite the higher convective variability typical of summer conditions.

Panel~\ref{fig:forecast_accuracy}c quantifies forecast skill by the annual relative error of daily peak SSRD at 12:00 local time across China. Errors are kept below ~20\% over most regions even at a 7-day lead. Their largest values (>30\%) are concentrated in southern and coastal areas where cloud dynamics are particularly hard to predict. The spatial distribution of errors indicates the most reliable performance of SunCastNet across northern and inland regions in which irradiation variability is impacted more by synoptic-scale dynamics than by local convection.


Figure~\ref{fig:gfs_comparison} compares forecasts of SSRD from SunCastNet (0.05°, 10-min) and the GFS~\cite{kalnay1996ncep,wu2023gfs} (0.25°, 1-hour, SSRD converted from downward shortwave radiation flux, DSWRF) against observations at 2,164 stations across China. Panel~\ref{fig:gfs_comparison}a shows that forecast errors, when averaged over all stations, increase with lead time for both models. The median relative error of SunCastNet rises from about 13\% at a 2-day lead to ~20\% at 7 days, while GFS errors grow from ~22\% to 28\% oaer the same horizon. SunCastNet maintains a ~5–10\% lower error than GFS across all lead times. Its $ 50\pm 25\%$ interquartile ranges are also consistently narrower (e.g., ~7–24\% at 2 days and ~9–37\% at 7 days) compared with GFS (16–33\% at 2 days and ~18–50\% at 7 days), indicating more stable performance across stations. 

We have conducted detailed assessments of data generated by SunCastNet and GFS between August 2020 and August 2025 over three representative subregions of China (Fig.~S1; background field in Data Supplementary). Across Northeast, Southeast, and Southwest China, SunCastNet consistently outperforms GFS under the three meterological conditions traditionally considered most challenging for solar forecasting: aerosol perturbations from straw burning \cite{li_reduction_2017} (e.g., in Feb 2020), typhoon structures unresolved at 25 km \cite{ceferino2022stochastic} (e.g., Typhoon Hagupit in July 2020), and frequent temperature inversions and basin effects \cite{jerez_impact_2015}. Notably, SunCastNet’s ModAFNO–diagnostic module appears to leverage first-frame solar radiation to infer related processes implicitly represented in the inputs (e.g., aerosols, cloud microphysics), enabling more accurate judgments than GFS whose radiation diagnostics don't accommodate such factors.

Panel~\ref{fig:gfs_comparison}b presents a representative rainy spell at one station in early April 2023. From the daily maximum irradiance, both models capture the sharp drops in irradiance associated with rainfall events. However, we observe that their temporal structures diverge in most circumstances. Specifically, SunCastNet has hour-to-hour fluctuations, whereas GFS tends to produce stereotyped triangular diurnal cycles, rising after forecast issuance, peaking at noon, and then declining in the afternoon. Furthermore, GFS tends to over estimate the irradiance in the afternoon every sunny day (\circled{\small 1}, \circled{\small 2} and \circled{\small 3} in Panel~\ref{fig:gfs_comparison}b; see also additional stations and cases in Data Supplementary). Plausibly, GFS contains limited hour-level information, given that the underlying moisture and cloud diagnostics are only updated at quasi-static 6-hour cycles \cite{NOAA_GFS_doc} with the Rapid Radiative Transfer Model (RRTM)~\cite{clough2005atmospheric}. As a result, the nominal hourly resolution of GFS may carry limited new information than 6-hour data, while SunCastNet explores finer-scale temporal data-embedded dependencies. We could expect less information contained in the GFS SSRD forecast than in SunCastNet.

To further probe into this hypothesis, panel~\ref{fig:gfs_comparison}c evaluates the forecasts using mutual information (MI). Unlike standard error metrics such as mean square error (MSE) which quantifies pointwise deviations between forecasts and observations, MI evaluates how much of the underlying temporal structure of the true series is preserved in the forecasts. At 97\% of the 2,164 stations, SunCastNet yields higher MI than GFS, with an average ratio of ~1.2 (about 20\% more shared information with the ground truth). This demonstrates that SunCastNet reduces forecast errors and keeps richer temporal and multivariate dependencies. Higher MI thus implies that SunCastNet provides forecasts with greater decision-making information for downstream tasks such as renewable-energy scheduling and operational management.

\subsection*{Decision-making under SunCastNet}

Based on the information-oriented comparison in Fig.~\ref{fig:gfs_comparison}, we next examine how differences in predictive content translate into decision-making outcomes. If SunCastNet indeed provides richer and more consistent signals than GFS, we expect such advantages to be available in both short-term operational scheduling and long-term investment strategies.

\begin{figure*}[htbp]
  \centering
  \includegraphics[width=0.8\textwidth]{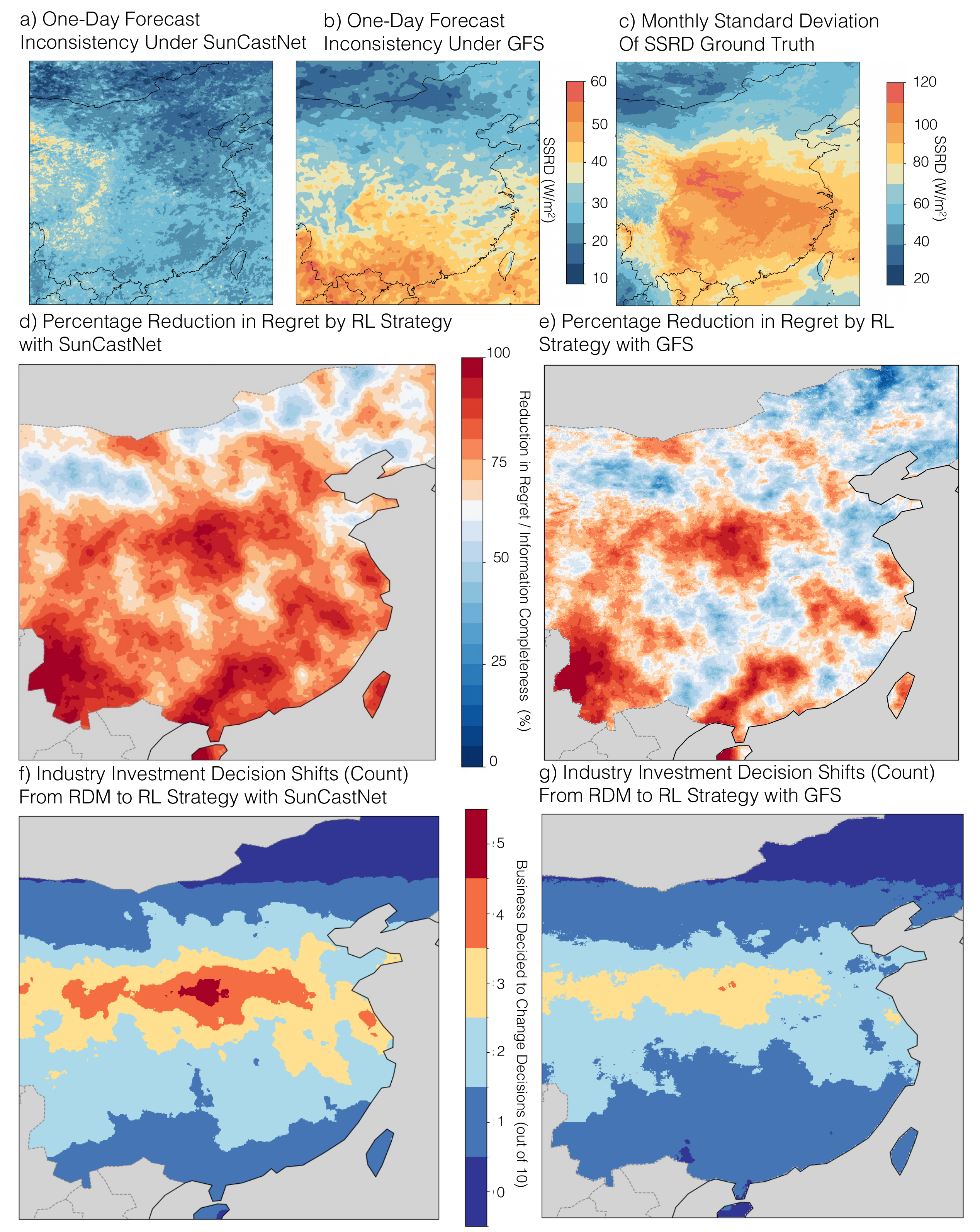}
  \caption{\textbf{From forecast consistency to decision outcomes under SunCastNet and GFS.} 
  (a–b) One-day forecast inconsistency for SunCastNet and GFS, defined as the difference between the predicted irradiation for day~2 issued on day~1 versus that issued on day~2 itself (red indicates larger inconsistency, blue smaller).  
  (c) Monthly standard deviation of SSRD from satellite ground truth, used as a naïve predictability baseline (red higher variability, blue lower).  
  (d–e) Percentage reduction in decision regret achieved by reinforcement learning (RL) battery management strategies trained with SunCastNet versus GFS forecasts, averaged over 10 representative industrial sectors (automobile, electronics, food processing, textiles, pharmaceuticals, chemicals, steel, paper, cement, and glass).  
  (f–g) Investment decision shifts (out of 10 sectors) when switching from a minimax robust decision-making (RDM) baseline to RL-informed strategies using SunCastNet or GFS forecasts (red more shifts, blue fewer; range 1–5).}
  \label{fig:rl_decisions}
\end{figure*}

We first define inconsistency as the difference between the predicted irradiation for day~2 issued on day~1 and that issued on day~2 itself. In Panels~\ref{fig:rl_decisions}a–b, SunCastNet exhibits much smaller inconsistencies than GFS: errors remain below 20~W\,m$^{-2}$ in most regions, rarely exceeding 30~W\,m$^{-2}$, whereas GFS shows widespread inconsistencies of 30–50~W\,m$^{-2}$ and hotspots above 50~W\,m$^{-2}$ in the Yangtze basin and Sichuan. Panel~\ref{fig:rl_decisions}c presents a baseline obtained by sampling 30-day ground truth irradiation sequences and reporting the standard deviation. This baseline often exceeds 100~W\,m$^{-2}$ in cloudy regions, indicating the intrinsic variability of the system. Both SunCastNet and GFS therefore provide more stable, informed guidance to decision makers, but SunCastNet is markedly more coherent across time.

When forecasts are inconsistent, optimal action recommended by one forecast may contradict those based on the next forecast. This leads to unstable operational strategies, high adjustment costs, and in practice discourages operators from using forecasts at all~\cite{genov2024balancing}. Instead, they may resort to conservative robust optimization which ignores most forecast information but avoids substantial financial losses.

Panels~\ref{fig:rl_decisions}d–e quantify the operational benefits of different temporal forecasts with RL battery management. We evaluate electricity demand from ten representative industrial sectors. Different forecasts are used to train RL policies, and regret is measured against the perfect-information benchmark. Compared to the RDM baseline derived from 30-day historical scenarios, SunCastNet-informed RL policies achieve 76–93\% regret reduction ($50\%\pm25\%$ quantiles) across most of northern and eastern China, approaching 100\% in some regions. GFS-based RL policies reduce the regret by 43–66\% ($50\%\pm25\%$ quantiles). This demonstrates that higher information content and temporal coherence directly enhance short-term operational efficiency. As shown in Fig. S2, these improvements are robust across all ten representative industrial sectors. These sectors vary in spatial patterns but consistently show substantial efficiency gains under SunCastNet-informed RL.

Panels~\ref{fig:rl_decisions}f–g extend the analysis to long-term investment outcomes. Using ERA5-driven retrospective forecasts spanning over 25 years, we identify  industry-specific solar–battery projects that exceed the commercial viability  threshold of $\text{IRR}>12\%$. Under SunCastNet-informed RL, many regions with high irradiation variability (e.g., central China provinces such as Henan, Hubei and Anhui) shift their attitudes: they used to consider these projects "infeasible" under RDM but now deem them as "profitable" proposals under SunCastNet-informed RL. Panels~\ref{fig:rl_decisions}f–g show that, up to five out of ten sectors per region change their investment decisions, compared with usually two or three under GFS. This indicates that improved forecasts support better day-to-day scheduling and enlarge the set of industrial actors and geographies potentially attracted by solar investments, offering practical pathways to accelerate carbon-neutral industrial transitions.

While the comparison between SunCastNet and GFS demonstrates the benefits of higher spatial and temporal resolution,  the forecast horizon operates as another equally important dimension for both numerical and AI-based weather prediction. As shown in Fig.~S3, the economic benefit of solar forecasts depends strongly on horizon length. With only 1–2 days of lookahead, regret reductions remain modest (typically below 40\%), offering limited support for industrial scheduling. In contrast, horizons of 3–5 days yield meaningful gains, with regret reductions rising to 40–60\%. The benefits become substantial at the 7-day horizon, exceeding 70\% in many regions. According to these results, longer-horizon forecasts are essential to industrial solar–battery systems, since time-of-use(TOU)-based “valley charging–peak discharging” strategies require several days of advance planning to capture their full economic value.

\section*{Discussion}

We have developed SunCastNet, a sequential AI framework that generates high-resolution (0.05°, 10-minute) solar radiation forecasts up to 7 days ahead by integrating specialized neural operators and diffusion model. When validated against 2,164 meteorological stations across China, SunCastNet output 5-10\% lower relative errors and 20\% higher mutual information compared to GFS, while maintaining superior temporal consistency. By integrating these forecasts with reinforcement learning for battery management, we present that improved solar predictions translate directly into economic value: reducing operational regret by 70-90\% and enabling up to five of ten industrial sectors per region to exceed the 12\% IRR viability threshold for solar investments. This work provides quantitative evidence that AI-driven improvements in solar forecasting can accelerate industrial decarbonization by making solar-battery systems economically viable across broader geographic and sectoral domains.

To ensure that the 25-year SunCastNet-powered retrospective experiments (August 2000 to August 2025) are not biased by potential overfitting to SunCastNet's training period (2015–2020), we conduct an additional robustness check. Specifically, we repeat the planning experiments five times using only data between August 2020 and August 2025 for both SunCastNet and GFS forecasts. As shown in Fig.~S4, although the spatial patterns of regret reduction differ slightly from the 25-year experiment, SunCastNet consistently delivers markedly higher regret reduction than GFS across most regions in China. It is worth noting that modern deep learning–based weather forecasting systems are trained on very large multi-decadal reanalyses, which decreases the likelihood of conventional overfitting~\cite{pathak2022fourcastnet,keisler2022forecasting}. The robustness test therefore showcases that the performance gains don't result from longer hindcast availability or overlap with the training set, but instead from real improvements in forecast quality.

Two limitations should be noted. First, our evaluation is confined to China, and applications to regions with sparser observations or different climatic regimes will require further validation. Second, while RL-based metrics highlight substantial economic benefits, real-world feasibility depends on regulatory structures. In the Chinese context, most industrial consumers currently are not allowed to sell electricity back to the grid, which constrains potential returns. 

Finally, we advocate that forecast evaluation for energy applications should move beyond RMSE. Measures of information content, temporal consistency, and regret more directly link forecast skill to economic outcomes. As solar capacity scales up, such evaluation will be crucial for both model development and planning.

\renewcommand\refname{References and Notes}
\bibliography{sample}

\paragraph{Competing interests:} Ten co-authors (Z.D., B.B., A.C., J.L., T.K., J.S., R.C., Y.Zh., R.Zh. and F.H.) are employees of NVIDIA Corporation, a commercial entity that develops and markets computing hardware and software products. All other authors are affiliated with academic or research institutions. The computational resources used in this study, including GPUs for model training and inference, are products manufactured by NVIDIA. The methods and results presented in this paper may be incorporated into future commercial products and services offered by NVIDIA. 

\section*{Data and materials availability:}
Most of the data used to train and evaluate our forecasting pipeline can be obtained from publicly available sources. The ERA5 dataset is available from the Climate Data Store (CDS) (\url{https://cds.climate.copernicus.eu}). The Global Forecast System (GFS) forecasts are provided by the National Centers for Environmental Prediction (NCEP) and can be accessed through the NOAA NOMADS system (\url{https://nomads.ncep.noaa.gov}) or via the NCAR Research Data Archive (\url{https://rda.ucar.edu}). The East Asia–Pacific Longwave/Shortwave Downward Radiation dataset (2016–2020) is available from the National Tibetan Plateau Data Center (\url{https://data.tpdc.ac.cn} )~\cite{letu2022new}. Ground-based solar radiation measurements are obtained from the China Meteorological Administration (CMA) via the TianQing Data Service System, upon request through the National Meteorological Information Center (\url{http://data.cma.cn} ). All plots were made using Matplotlib, Xarray, and NumPy, and geographical maps were produced using Cartopy. The 42 industrial cases analyzed in this study are derived from real operational data that the authors accessed directly through industrial collaborations. These datasets reflect actual production and electricity demand patterns in representative enterprises, ensuring practical relevance. However, due to commercial confidentiality and non-disclosure agreements with industry partners, the raw data cannot be made publicly available.

Our CorrDiffSolar code is available at \url{https://github.com/NVIDIA/physicsnemo}. 
The complete SunCastNet framework(SFNO, ModAFNO, solar radiation diagnostics, and CorrDiffSolar) is provided at \url{https://github.com/NVIDIA/earth2studio}, 
and executable notebooks for reproducing our experiments are available at \url{https://github.com/kelvinfkr/Solar_Economy}. Sample output data (higher resolution comparing to Data Supplementary) with also Codebook to guide the inference of SunCastNet is available at \url{https://doi.org/10.6084/m9.figshare.30070609}. 

\subsection*{Supplementary materials}
Materials and Methods\\
Supplementary Text\\
Figs. S1 to S4\\
Tables S1 to S2\\
References \textit{(61-80)}\\ 
Code S1\\
Data S1

\clearpage
\setcounter{page}{1}
\renewcommand{\thepage}{S\arabic{page}}

\fancyfoot{}  
\fancyfoot[R]{\small\sffamily\bfseries\thepage}  
\renewcommand{\thefigure}{S\arabic{figure}}
\setcounter{figure}{0}  

\renewcommand{\thetable}{S\arabic{table}}
\setcounter{table}{0}   

\renewcommand{\theequation}{S\arabic{equation}}
\setcounter{equation}{0}

\end{document}